\documentclass[prb,twocolumn,showpacs,amsmath,amssymb]{revtex4}
\usepackage{dcolumn}
\usepackage{graphicx}
\usepackage{bm}
\begin{document}
\title{Nonequilibrium interacting electrons in a
ferromagnet}
\author{P. F. Farinas}
\affiliation{Instituto de F\' \i sica, Universidade Federal do Rio de Janeiro,
21645-970, Rio de Janeiro, RJ, Brazil}

\date{\today}

\begin{abstract}
Dynamics of the magnetization in
ferromagnets is examined in the presence of transport electrons
allowing the latter to interact.
It is found that
the existence of inhomogeneities such as
domain wall (DW) structures,
leads to
changes that affect the
dynamical structure of the equations of motion for
the magnetization.
Only in the limit of
uniform magnetizations or
sufficiently wide DW's,
the equations of motion maintain the form
they have in the noninteracting case. In this
limit, results like
the spin torques, the Gilbert
parameter, and the DW velocities become renormalized.
However the length scale that defines such a limit
depends on the strength of the interaction.
It is shown that if
large ferromagnetic fluctuations exist in
the metallic band then the range for which
conformity with the noninteracting case holds
extends to the limit of arbitrarily narrow DW's.
%and theoretical estimates of the
%DW terminal velocity made in the abscence of such interelectronic
%interactions can be corrected to better agree with observed
%values.

\end{abstract}
\maketitle

%\narrowtext

Increasing possibilities for spin selective transport of electrons
in different materials
and their promising physical and technological consequences
have yielded a broad variety of research on electronic
motion in the presence of a ferromagnetic background.
Interest ranges from semiconductors
to transition metal
ferromagnets\cite{ohno,vig,ticoo,wolf,sarna,kohda,bailey}.

A question under current attention is how electrons
flowing in a
ferromagnetic media should modify the dynamics of its order
parameter, ${\rm \bf M}({\rm \bf r}, t)$.
In the recent years, a typical route to handle this problem
has consisted of treating conduction
and valence electrons distinctively, the former
assumed to belong to a $s$-band, while the latter being
in a filled $d$-band responsible for the local moments
that give rise to
${\rm \bf M}$.

One often uses some equation of motion
(as, e.g., the Landau-Lifschitz-Gilbert (LLG)\cite{llg})
to treat ${\rm \bf M}$ as a classical
field, while a 
kinetic transport equation
is used for the dynamics of the conduction
$s$-electrons, usually in the limit
of long length scales where semiclassical equations
are valid for the electronic spin\cite{silin}.
Coupling of these equations is 
achieved by including
a $s$-$d$ channel through
a Kasuya interaction operator\cite{kasuya,larsen},
whose form is
familiar, {\it e.g.},
in the Kondo physics of dilute magnetic alloys\cite{kondo}.
In transition
metal ferromagnets this approach has been used
in
attempts to draw an
adequate picture to understand
experiments\cite{e1,e2,e3,e4}.
Recently discussed issues include the
role of
the so called nonadiabatic spin torque
that the electronic current produces on a
domain wall (DW), affecting its motion and stability,
and the contribution of the conduction electrons for the
damping of ${\rm \bf M}$\cite{zhang1,zhang2}.

Underlying the theories used to treat these phenomena we may identify
two assumptions (among many others). First,
that the ferromagnetic state is a broken symmetry
of the interacting electrons within the $d$-band. Second, that
the $s$-band interelectronic interaction ($s$-$s$) leads only to trivial
renormalizations or else that the $s$-$s$ interaction is sufficiently weak
not to produce any measurable effects on ${\rm \bf M}$. Thus while the
interaction is considered to be sufficiently important in the $d$-band
to drive its electrons into a ferromagnetic state, the $s$-band is
taken as an essentially noninteracting-electrons' metallic band.

In this article we relax the
second of these assumptions and investigate some preliminary general consequences
that are expected to be seen on the dynamics of
the magnetization. As we will see,
the interactions may affect the dynamical structure of the equations
of motion. In the limit
of uniform magnetizations or sufficiently wide DW's, treating
the $s$ electrons as noninteracting is justified with the
proper renormalizations.
We show
that in the particular limit for which
large ferromagnetic fluctuations exist in the
$s$-band, decoupling
of the equations of motion in a local frame is possible for
narrow DW's. Consequently, not only the length for which the
``wide DW'' approximations become valid is lowered, but
the equations of motion acquire the form they
have for noninteracting
$s$-electrons.
%with renormalized coefficients that can
%be adusted to imrove agreement with experiments
%for early estimates of the DW terminal
%velocity\cite{zhang1}.

%In the language of the renormalization group,
%the
%interactions become irrelevant as the system
%of $s$-electrons approach a ferromagnetic
%transition.

For these
purposes, we take the usual algorithm described in the third paragraph,
and modify it to include the $s$-$s$ interactions. This is done
by using the Fermi liquid theory\cite{silin,landau,baym} and
the results should hold provided the conduction
electrons are normal (no assumption
is required to whether the interactions are weak or not).

The transport equation for the $s$-electrons'
distribution can be written in a
compact form\cite{silin},
\begin{equation}
\partial_t{\cal N}
+\left\{
{\cal N},{\cal H}\right\}
+\frac{i}{\hbar }\left[
{\cal H},
{\cal N}\right]
=I[{\cal N}],
\label{wig}
\end{equation}
where all quantities are $2\times 2$ matrices in spin space,
${\cal N}_{\rm \bf p}({\rm \bf r},t)$ is the full distribution function
for the $s$-electrons,
${\cal H}_{\rm \bf p}({\rm \bf r},t)$ is the effective hamiltonian
and
$I[{\cal N}]$ is a collision integral. Curly Poisson and
straight commutator brackets are also implied in the
notation. The commutator is kept with respect
to the spin matrices as a consequence of using a local quantization
axis. These functions must be understood as averages in the
state of the system, which satisfy quantum mechanical
equations of motion that reduce to the above semiclassical
equation in the long wavelength limit\cite{kadanoff}.

Inclusion of the $s$-$d$ coupling to the order
parameter of the ferromagnetic band is achived by using
an exchange interaction besides the Fermi liquid
hamilitonain ${\cal H}_{FL}$,
$
{\cal H}={\cal H}_{FL}+{\cal H}_{sd},
$
where
$%\begin{equation}
{\cal H}_{sd}=SJ_{ex}{\bm \sigma }\cdot
{\rm \bf M}({\rm \bf r},t)/M_s,
%\label{exch}
$%\end{equation}
$S$ is the projection of the local spin for the $d$-electrons,
$M_s$ is the saturation magnetization, and
${\bm \sigma }$ are the usual Pauli matrices\cite{zhang1}. Note that
the structure of the transport equation in spin space carries
a commutator even in the absence of the $s$-$d$ exchange coupling,
which has long been known\cite{silin}, however the intrinsic
$s$-electron's commutator vanish if these electrons are noninteracting.

To write equations of motion from the transport equation, we
trace Eq.(\ref{wig}) to obtain equations for the local
spin density of the $s$-electrons ${\rm \bf m}= {\rm tr}[{\bm \sigma }
{\cal N}]$ and the spin current tensor
${\bm J}_i \equiv {\rm tr}[{\bm \sigma }
(\partial {\cal H}/\partial p_i){\cal N}]$.

Before
writing these equations we separate, as in Ref.\cite{zhang1},
the local spin density in terms of an adiabatic
component in the direction of ${\rm \bf M}$ and a transverse
(nonadiabatic) component ${\delta {\rm \bf m}}$,
\begin{equation}
{\rm \bf m}=\frac{n_0'}{M_s}{\rm \bf M}+{\delta {\rm \bf m}},
\label{mag1}
\end{equation}
where $n_0'=n_0/(1+F_0^a)$, $n_0$ is the local equilibrium
spin density, and $F_{\ell}^a$ is the usual notation for
the standard Landau antisymmetric dimensionless
parameters.

The spin current tensor can similarly be split in two contributions,
\begin{equation}
{\bm J}_i = -\frac{\mu_BP'}{eM_s}{j_{e}}_i{\rm \bf M}
+{{\bm j}_{\sigma }}_i,
\label{spcurr}
\end{equation}
where ${\bm j}_e$ is the electric current density,
$e$ is the electron's charge, $\mu_B$ is the Bohr magneton,
and $P'=P/(1+F_0^a)$ with the current's spin polarization
given by $P$.

We note that the adiabatic responses to
${\rm \bf M}$ are renormalized by the
interaction in exactly
the same way as the Pauli susceptibility in a paramagnet.
This comes from the $s$-$d$ coupling that effectively
adds a local contribution
\begin{equation}
{\rm \bf h}_{\rm \bf p}^0 = \frac{SJ_{ex} }{2M_s}{\rm \bf M},
\label{mol}
\end{equation}
to the
molecular field ${\rm \bf h}_{\rm \bf p}=
{\rm \bf h}_{\rm \bf p}^0 + {{\rm \bf h}_{\rm \bf p}}_{FL}$
acting on the Fermi liquid quasiparticles\cite{baym}. Formally,
the local field ${\bm S}=-S{\rm \bf M}/M_s$ plays the
role of a ``magnetic field'' in the equations of motion
of a uncharged Fermi liquid.

The equation of motion for the local spin density becomes
\begin{equation}
\frac{\partial {\rm \bf m}}{\partial t}+\nabla_i{\bm J}_i =
-\frac{1}{\tau_{ex}M_s} {\rm \bf M}\times \delta {\rm \bf m}
-\frac{\delta {\rm \bf m}}{\tau_{sf}},
\label{mag}
\end{equation}
where sum is implied over repeated indices.
Equation (\ref{mag}) is the continuity equation
for the $s$-spins modified by keeping
the part of the collision integral that does not
conserve spin (here we use a relaxation time 
approximation with $\tau_{sf}$ for
scattering processes other
than the ones mediated by the $s$-$s$ interaction). The term with
$\tau_{ex}=\hbar/SJ_{ex}$
comes from the $s$-$d$  hamiltonian that affects the molecular
field through ${\rm \bf h}_{\rm \bf p}^0$ given by Eq.(\ref{mol}).

For the spin current, we obtain,
\begin{eqnarray}
&&\partial_t {{\bm j}_{\sigma }}_i+c_s^2\nabla_i\delta {\rm \bf m}
\nonumber \\
&&=\left[\frac{\hbar {\rm \bf M}}{\tau_{ex}M_s}
+\frac{2{\rm \bf m}}{\hbar N(0)}\left(F_0^a-\frac{F_1^a}{3}\right)\right]
\times {{\bm j}_{\sigma }}_i
-\frac{{{\bm j}_{\sigma }}_i}{\tau'_{sf}},
\label{cur}
\end{eqnarray}
with
$c_s^2 = (v_F^2/3)(1+F_0^a)(1+F_1^a/3)$, $N(0)$ the
density of states at the $s$-band unpolarized Fermi surface,
and ${\tau'_{sf}}^{-1}=(1+F_1^a/3){\tau_{FL}}^{-1}+{\tau_{sf}}^{-1}$.
The fact that ${\tau'_{sf}}\ne {\tau_{sf}}$ results from the internal
spin diffusion in the Fermi liquid, for which spin-current is not
a conserved quantity.
Such a difference
exists even in the abscence of interaction (as is easily
seen by putting $F_{\ell }=0$) although this has not been always considered.

Now, with the usual assumption that the $s$-electron's dynamics is much
faster than that of ${\rm \bf M}$, we look for the solution
of Eq.(\ref{cur}) after the transient part has practically
vanished out, what occurs after a few relaxation times. At
these times, the quasi-steady solution consists of the
precession of the spin current about ${\rm \bf M}$, so that
the Eq.(\ref{cur}) becomes
\begin{equation}
g{\rm \bf m}
\times {{\bm j}_{\sigma }}_i
+{{\bm j}_{\sigma }}_i
=-D_s\nabla_i\delta {\rm \bf m},
\label{cur2}
\end{equation}
where $g=2\tau'_{sf}(F_0^a-F_1^a/3)/\hbar N(0)$ measures
the strenght of the interaction and $D_s=v_F^2\tau'_{sf}(1+F_0^a)/3$
is the spin diffusion coefficient.

In Eq.(\ref{cur2})
the differences between interacting and noninteracting
$s$-electrons become apparent. By setting the interaction to
zero ($F_{\ell }=0$) one recovers Fick's relation
between spin current and spin density.
The additional term on the left side
originates from the precession of the $s$-electrons spin current
about the local spin density, the same phenomenon that 
causes the long
known Leggett-Rice effect in normal $^3$He at very low
temperatures\cite{leg,cor}.

Combining Eqs.(\ref{cur2}) and (\ref{mag}), we obtain
\begin{eqnarray}
D'_s\nabla^2 \delta {\rm \bf m}
-\frac{1}{\tau_{ex}M_s}\delta {\rm \bf m}\times {\rm \bf M}
-\frac{\delta {\rm \bf m}}{\tau_{sf}}+{\bm \Delta }\nonumber \\
=\frac{n'_0}{M_s}\frac{\partial {\rm \bf M}}{\partial t}
-\frac{\mu_BP'}{eM_s}({\bm j}_e\cdot {\bm \nabla }){\rm \bf M},
\label{delm}
\end{eqnarray}
where $D'_s=D_s/(1+g^2n'_0)$ and
\begin{equation}
-\frac{\bm \Delta }{D'_s}=
\frac{gn'_0}{M_s}\nabla_i({\rm \bf M}\times \nabla_i\delta {\rm \bf m})
+\left(\frac{gn'_0}{M_s}\right)^2
\nabla_i[{\rm \bf M}({\rm \bf M}\cdot \nabla_i\delta {\rm \bf m})].
\label{del}
\end{equation}
%We are keeping leading order in quantities like
%$\delta {\rm \bf m}$, ${\bm j}_e$,
%${{\bm j}_\sigma}$, and ignoring
%terms such as $\partial^2{\rm \bf M}/\partial t^2$.
%Hence we have dropped
%$\partial \delta {\rm \bf m}/\partial t$
%(it is clear from
%Eq.(\ref{mag1}) and the rotation invariance 
%${\rm \bf M}\cdot \partial {\rm \bf M}/\partial t=0$ that
%$\delta {\rm \bf m}\sim \partial {\rm \bf M}/\partial t$).
Equation (\ref{delm}) is the central result of this
article. Its form suggests that
since the interaction between $s$-electrons
is in general not weak, the
use of an independent-electrons's approach in the $s$-band,
as it has been common in the literature for $s$-$d$-coupling
models, is not correct in principle. Regardless
of the equation of motion satisfied by ${\rm \bf M}$,
one sees by a careful examination of
Eqs.(\ref{delm}) and (\ref{del}) that the
dynamics obtained for ${\rm \bf M}$
in the absence of $s$-$s$ interaction
will be in general rather different from that in its
presence.
%The first question, however, and the one to be addressed
%in what follows,
%is on the existence or not of limits for which the effects of the
%$s$-$s$ interaction reduce to simple renormalizations and the
%FEA becomes admissible.

To illustrate how the $s$-$s$ interaction may affect the
dynamics of ${\rm \bf M}$, we shortly indicate how
the result given by Eq.(\ref{delm}) changes
the widely used LLG equation\cite{llg},
without questioning its limits of validity\cite{mac},
since this specific choice does
not change the general conclusions we want to draw in
what follows.

Interaction of the spin density $\delta {\rm \bf m}$
with ${\rm \bf M}$, brings an
additional torque to the LLG equation,
%\begin{equation}
%\frac{\partial {\rm \bf M}}{\partial t}=
%-\gamma {\rm \bf M}\times {\rm \bf H}_{eff}
%+\frac{\alpha }{M_s}{\rm \bf M}\times
%\frac{\partial {\rm \bf M}}{\partial t}
%+ {\rm \bf T},
%\label{llg}
%\end{equation}
%where
${\rm \bf T} = (1/\tau_{ex}M_s)\delta{\rm \bf m}
\times {\rm \bf M}$.
From this torque $\delta {\rm \bf m}$ can be
taken as a function ${\rm \bf M}$ and
used in Eq.(\ref{delm}) to give an
equation for the dynamics of the order
parameter ${\rm \bf M}$. A more detailed calculation
of how the dynamics of ${\rm \bf M}$ changes with the
$s$-$s$ interaction will be shown elsewhere, but it is
clear, by simple inspection of Eqs.(\ref{delm}) and (\ref{del})
that the result obtained will be two quite different
equations for interacting and non-interacting $s$-electrons.

In the limit of vanishing $s$-$s$ interaction these
expressions give the known results for the equations
of motion\cite{zhang1}. The additional dynamical
structure in the interacting case comes
from the term ${\bm \Delta }$ given in Eq.(\ref{del}).
Besides this
additional contribution, there are also renormalizations
of the coefficients in the other terms. We emphasize that
our result is exact, apart from the usual assumptions
of Fermi liquid theory and the general hypotheses that
are used to validate the $s$-$d$ model.

We see that in the limit
of strong interactions ${\bm \Delta }$ may introduce changes
that lead to different dynamics. As a matter of fact,
the strongly interacting limit is somewhat difficult to
evaluate, since while $n'_0=n_0/(1+F_0^a)$ becomes negligible
in the limit of large $F^a_{\ell }$'s,
the exact behavior of the product
$gn'_0$, that governs ${\bm \Delta }$ is
difficult to predict without an explicit knowledge
of the Landau parameters and also the effective mass
that enters $N(0)$. One can, nonetheless, keep
the values of the Landau parameters finite and $\ne 0$
to study the
overall differences that ${\bm \Delta }$ is expected to
produce.

It is immediately noted,
however, that ${\bm \Delta }$ is only different from zero if the
gradients of ${\rm \bf M}$ are not zero, i.e., in the presence
of inhomogeneities. For uniform or nearly uniform magnetizations,
one sees that the sole effect of the interaction is to renormalize
the equations of motion through the coefficients $P'$ and $n'_0$.

In what follows, we will avoid the more complicated task of solving the full
equation for a finite ${\bm \Delta }$ (Eq.(\ref{delm})), and rather
concentrate in presenting a precise meaning for the
limit of nearly uniform
magnetization when $s$-$s$ interactions are not zero.

First we assume that in this limit, the DW's width
(or, equivalently, the characteristic average width of the
inhomogeneity) is
very large compared to any diffusion length of the
problem, so that ${\rm \bf M}$ varies slowly in space.
Under this assumption, decoupling of Eq.(\ref{delm})
is possible in a local frame in a similar way that it is
accomplished in
the noninteracting case. Taking the $z$ axis of such a frame parallel to
${\rm \bf M}$ yields the following equation,
\begin{equation}
{\bm \nabla }^2\delta {\rm m}_--\frac{1}{\lambda^2}\delta {\rm m}_- = f_-({\rm \bf r}),
\label{dec}
\end{equation}
where $\delta {\rm m}_-=\delta {\rm m}_x -i \delta {\rm m}_y$, $f_-=f_x-if_y$,
\begin{equation}
{\bm f}({\rm \bf r})=\frac{1}{1+ign'_0}\left[
\frac{n'_0}{M_s}\frac{\partial {\rm \bf M}}{\partial t}
-\frac{\mu_BP'}{eM_s}({\bm j}_e\cdot {\bm \nabla }){\rm \bf M}
\right],
\label{fr}
\end{equation}
and $\lambda^2=D'_s(1+ign'_0)/(\tau_{sf}^{-1}+i\tau_{ex}^{-1})$.
Equation (\ref{dec}) admits a standard Green's function solution,
\begin{equation}
\delta {\rm m}_-=\frac{1}{4\pi }\int d{\rm \bf r}'
\frac{e^{-|{\rm \bf r}-{\rm \bf r}'|/\lambda }}{|{\rm \bf r}-{\rm \bf r}'|}
f_-({\rm \bf r}'),
\label{sol}
\end{equation}
Let the typical scale over which ${\bm f}({\rm \bf r})$ varies be $w$.
It is clear from Eq.(\ref{fr}) that $w$ is given by the average
distance over which ${\rm \bf M}$ varies in space. For DW structures,
$w$ is given by the average DW's width.
If
$
w>>|\lambda |
$
then we can take $f_-({\rm \bf r})$ out of the integral in Eq.(\ref{sol}),
to obtain
$
-\delta {\rm m}_-/\lambda^2=f_-({\rm \bf r}),
$
and, from Eq.(\ref{dec}), ${\bm \nabla }^2 \delta {\rm \bf m} = 0$
and ${\bm \Delta }=0$.

Equation (\ref{delm}) then aquires
the same form as the one used for noninteracting
$s$-electrons,
and the torque and other quantities
can be determined by simply replacing the renormalized
parameters in the results earlier obtained in the
abscence of $s$-$s$ interaction\cite{zhang1,zhang2}.

The condition $w>>|\lambda |$ that
validates this ``wide DW'' approximation
can be explicitly written as
\begin{equation}
w>>\sqrt{\frac{v_F^2\tau'_{sf}\tau (1+F_0^a)}{3\sqrt{1+(g{n'_0})^2}}}
=|\lambda |,
\label{cond}
\end{equation}
where
$\tau^{-2}=\tau_{sf}^{-2}+\tau_{ex}^{-2}$.

We see that the characteristic length for this
limit to be set depends on the strength
of the interaction. Again, the simple limit of very large
interactions is difficult to evaluate since one does not
in general know
how the Landau parameters and the effective mass are
exactly related.

However, other strongly-correlated limits
are approached when the $s$-electrons are driven towards
some instability. The general Pomeranchuk instability\cite{baym}
is set when $F^a_{\ell }/(2\ell +1)\rightarrow -1_+$. There are
other correlation driven instabilities whose
onsets could be studied, like the
metal-insulator transition or the SDW instability\cite{prl}.

We will focus here on the Stoner instability
that is approached when $F_0^a\rightarrow -1_+$.
The onset of such an instability is characterized by
the presence of large ferromagnetic fluctuations
in the system of $s$-electrons.
If one thinks of a ``weakly interacting'' $s$-band
it may appear unphysical to assume the $s$-electrons
are in such a regime. However, we just showed
that no physical ``weakly interacting'' scenario
exists in a non-trivial solution ($F_{\ell }$'s not all
equal to zero). The electrons
are rigorously undistinguishable, $s$-$d$ models
separate them in two groups with hopes that this will
be an effective descripition. It is quite compelling then, that
the strong fluctuations that are present in the $d$-electrons'
part of the wave function, will be also shown in the $s$-electrons'
part.
Such a regime is analogous to
the so-called ``nearly ferromanetic
Fermi liquid'', a regime that, {\it e.g},
normal $^3$He undergoes as its temperature is lowered
at room pressure\cite{baym}.

In the presence of large ferromagnetic fluctuations,
Eq.(\ref{cond}) becomes
\begin{equation}
w>>\sqrt{\frac{v_F^2\tau'_{sf}\tau }{3gn_0}}(1+F_0^a)
\rightarrow 0,
\label{cond2}
\end{equation}
and the effective length defining the ``wide DW'' limit
is arbitrarily lowered. It is clear, however,
that a lower limit exists for
the usual approximations to be valid (for example, the
semiclassical equation of motion is valid for long wavelengths),
however these scales are usually much less than the typical
diffusion paths that relate to $|\lambda |$ in the limit
of vanishing $s$-$s$ interaction.

As examples,
the renormalizations of the
gyromagnetic ratio and Gilbert
parameter in the LLG equation
in the regime of
strong ferromagnetic fluctuations, are given by
\begin{equation}
\gamma' =
r\gamma ,
\;\;\;{\rm and}\;\;\;
\alpha'=r\alpha
+\frac{\tau_{ex}}{\tau_{sf}}.
\label{gyr}
\end{equation}
where $r = M_s(1+F_0^a)(1+\tau^2_{ex}/\tau^2_{sf})/n_0$.

In closing, we considered the effect of the
interelectronic ($s$-$s$) interaction in a $s$-band
coupled to a ferromagnetic $d$ band, and found
that structural changes in the dynamics of the
order parameter ${\rm \bf M}$ may occur.

The
theory correctly reproduces the nointeracting
limit that
has been
used  as the standard approach for electrons
flowing in ferromagnetic metals, which include
transition metals and their alloys.
We saw that,
in the general case,
the only safe assumption to boldly ignore the $s$-$s$ interaction would be
that its effective amplitude is too small, which does not appear to be
a plausible demand for these systems.

In the particular regime of large ferromagnetic
fluctuations in the $s$-band, the inclusion
of the $s$-$s$ interaction is shown to
produce two main effects: 1 - rescaling the
equations of
motion, and 2 - extending the ``wide inhomogeneity''
limit to arbitrarily
narrow domain walls.

This latter effect may be physiscally understood by noticing that
in such a regime of large ferromagnetic fluctuations
the spin density (${\rm \bf m}$) in the $s$
band becomes highly susceptible to changes in the
magnetization (${\rm \bf M}$) (analogously to what happens in a
usual paramagnet at the onset of a ferromagnetic instability).
The closer the instability is,
the stronger the tendency of
${\rm \bf m}$ to track ${\rm \bf M}$,
so that the length scales within which ${\rm \bf M}$
changes must be narrower in order for the dynamics
of ${\rm \bf m}$ to be affected by them. This
is appreciated only if the interactions are properly
accounted for.

The fact that
early approaches that do not
consider the $s$-$s$ channel work reasonably well
in explaining
some observed properties\cite{zhang1,zhang2}, may
be suggestive that the $s$-band electrons
in the theoretical description of these materials
must be considered as a
strongly correlated state with large ferromagnetic
fluctuations.

The author aknowledges FAPERJ, Rio de Janeiro, Brazil,
for providing means for this work to be completed.

%Typical ferromagnets that are experimentally investigated
%include transition metals and their alloys. Estimates for
%the constants involved in the theory are $J_{ex}\sim $ eV,
%$\tau_{sf}\sim 10^{-12}$ s, $\tau{ex}\sim 10^{-14}$ s,
%$n_0/M_s\sim 10^{-2}$.

%Using these values in the expression
%for the renormalized terminal DW velocity in the abscence
%of an external magnetic field, $v_{T}=-c_J/\alpha' $,
%where $c_J$ is the coefficient of the nonadiabatic
%contribution to the spin torque we get 
%\begin{equation}
%F_0^a\sim -1 + \frac{10^{-2}}{\sqrt{\alpha }}\sqrt{\frac{v_T^0}{v_T}},
%\label{vel}
%\end{equation}
%where $v_T^0$ is the theoretical estimate made in the
%abscence of $s$-$s$ interaction and $v_T$ is the experimentally
%observed value. Assuming the bare Gilbert parameter to be
%in 
%$10^{-2}<\alpha <10^{-1}$

\end{document}